\documentclass[%
10pt,
nofootinbib,
nobibnotes,
aps, 
physrev,
]{revtex4-2}

\usepackage[caption=false]{subfig}
\usepackage{ragged2e}

\usepackage{graphicx}% Include figure files
\usepackage{dcolumn}% Align table columns on decimal point
\usepackage{bm}% bold math

\usepackage{hyperref}% add hypertext capabilities

\usepackage{amssymb}
\usepackage{amsmath} % needed for some maths symbols
\usepackage{braket} % needed for ket notation
\usepackage{xcolor}

\newcommand{\mono}[1]{\texttt{#1}}
\renewcommand{\vec}[1]{\textbf{#1}}
\newcommand{\fcaption}[1]{\caption{\small #1}}

\usepackage[textsize=footnotesize,backgroundcolor=red!25]{todonotes}

\begin{document}

\title{\textbf{Heuristics for multi-stage quantum walks to find Ising ground states} 
}% 

\author{Asa Hopkins}
\email{Contact author: asa.hopkins@strath.ac.uk}
\author{Viv Kendon}
\affiliation{%
Department of Physics and SUPA, University of Strathclyde, Glasgow G4 0NG, United Kingdom
}%

\date{\today}

\begin{abstract}
One way to approximate a quantum annealing schedule is to use multiple quantum walks chained together, without intermediate measurements, to produce a multi-stage quantum walk (MSQW). Previous work has shown that a MSQW is better than QAOA (quantum alternating operator ansatz) for solving optimization tasks using multiple stages [Gerblich et al, arXiv:2407.06663]. In this work, we develop an efficient heuristic for choosing the free parameters in MSQW, and use it to obtain improved scaling compared to single stage quantum walks. We show numerically that the heuristic works well for problems with a large minimum energy gap, giving a polynomial scaling in the number of stages, and leading to an overall algorithm that scales polynomially in time. For problems with a small minimum gap, the scaling breaks down such that adding more stages decreases the success probability, leading to an overall scaling that is exponential in time as expected. Our methods are general and can be applied to any optimization problem to obtain good annealing schedules.
\end{abstract}


\maketitle

\section{Introduction}

Many decision problems, for example Karp's 21 NP-complete problems \cite{Karp2}, can be encoded into an Ising Hamiltonian with a number of qubits which is polynomial in the problem size \cite{Karp}. The encoding is such that the solution to the problem corresponds to the Ising Hamiltonian ground state. By definition, all problems in NP can be reduced to an NP-complete problem in polynomial time, so all problems in NP can be encoded on a quantum computer in polynomial space and time. 
A quantum annealer is designed to instantiate such an Ising Hamiltonian directly, and then solve the problem by determining the ground state. Developing effective methods for finding the ground state is thus an important area of research. Physical quantum annealers usually apply a transverse field to provide dynamics, and vary the strength of the Ising Hamiltonian and transverse field to drive the system towards lower energies, and there is a large body of work exploring how best to do this, see, e.g., the review by \citeauthor{Crosson2020} \cite{Crosson2020}.

Our work in this paper on multi-stage quantum walks (MSQW) builds on recent work by \citeauthor{Callison} \cite{Callison,Perspectives} and \citeauthor{Lasse} \cite{Lasse}. In \cite{Callison} it is shown that single stage quantum walks can find ground states better than quantum search algorithms for spin glass problems, without requiring fine tuning of the hopping rate to achieve this. 
This improvement is possible by exploiting the correlations described by the coupling terms of the Ising model.  The expectation is, therefore, that quantum walks will perform well for a range of problems encoded into Ising Hamiltonians. \citeauthor{Mirkarimi2023} \cite{Mirkarimi2023} show numerically that single stage quantum walks perform well for instances of MAX 2-SAT that classical methods find hard, and vice versa, and suggest using them as part of portfolio solvers.
In \cite{Lasse} it is shown that a given number of quantum walk stages will always perform better than the same number of QAOA stages, whilst also being more robust to the choice of parameters. 

\citeauthor{Banks} \cite{Banks} study the behaviour of quantum walks for solving MAX-CUT problems, but focusing on the metric of expected energy rather than the probability of observing the ground state. Qualitatively, they show that multi-stage quantum walks are able to decrease the state energy, but also that each stage increases the state temperature when modelled as a thermal Gibbs state in the infinite time limit, implying the resulting state should be possible to simulate classically. However, this result seems dependent on the choice of Hamiltonian normalisation, as it is the quantity $\beta \hat{H}$ that defines the Gibbs state. Even so, this fits well with \citeauthor{Bode} \cite{Bode}, who replace quantum evolution in QAOA with a classical mean-field approximation, giving a classical algorithm to find a good approximation to the ground state in polynomial time, with a probabilistic bound placed on the error. It may be that such an approach could be used to replace quantum evolution with classical evolution in the multi-stage quantum walk algorithm, providing the efficient classical simulation predicted by \citeauthor{Banks}, up to the error bound. 

\citeauthor{Dennis} \cite{Dennis} show that it is possible to solve a spin-glass problem in polynomial time using a guided quantum walk, which is a multi-stage quantum walk with the parameters determined through classical optimization. However, their analysis then reveals that they have moved the exponential effort into the classical optimization in order to tune the walk parameters well enough. When they take into account both quantum and classical resources, then for an $n$-qubit problem they obtain an exponential scaling of $O(2^{0.12n})$ for solving exact cover problems with guided quantum walks, compared to $O(2^{0.52n})$ for a single stage quantum walk. It is interesting to note that, for this problem class, single quantum walks seem to perform no better than Grover's search, in contrast to the spin glass problems studied in \cite{Callison}.  Nonetheless, the spin glass problems are a useful test problem because they are uniformly hard, and thus show good scaling even for small problem sizes, and we use the available datasets \cite{data,TimData} for our numerical simulations.

In this work, we develop methods to specify the parameters in a MSQW using heuristics that are efficient to calculate and have a good theoretical basis. We begin in section \ref{sec:background} by defining the Ising Hamiltonian and the spin glass problems we use for numerical simulations. In section \ref{sec:methods} we present the theoretical tools we use to develop the heuristics. In section \ref{sec:results} we present our numerical tests of these heuristics, and in section \ref{sec:further} we summarise our findings and suggest further research directions.  Technical details for some of our results are in the appendices.

\section{Background}\label{sec:background}
For a system of $n$ spins, an Ising Hamiltonian can be described with a set of $n$ magnetic field values $h_i$, and a set of $\frac{n(n-1)}{2}$ interaction values $J_{i,j}$, where $j < i$. The Hamiltonian is then defined as
\begin{equation}\label{IsingJ}
\hat{H}_I = -\sum_{i = 0}^{n-1} h_i \hat{Z}_i - \sum_{i = 1}^{n-1} \sum_{j = 0}^{i-1} J_{i,j} \hat{Z}_i \hat{Z}_j.
\end{equation}
In this paper we map between spins and qubits such that $\hat{H}_I$ acts on a register of $n$ qubits, where the two available states are interpreted as the qubit having a spin of $+1$ and $-1$ respectively. The action of $\hat{Z}_i$ is then to multiply the state by the spin of qubit $i$. The solution to the original problem then corresponds to the ground state of this Ising Hamiltonian, which itself will be a particular assignment of spins to the qubits.

Many of the results in this paper hold for any Ising-encoded problem, but specific examples will focus on solving the Sherrington-Kirkpatrick (SK) spin glass problem. In this context, the SK spin glass problem can be defined as an Ising Hamiltonian where all $h_i = 0$, and all $J_{i,j}$ are sampled from a standard normal distribution. This system has a $\mathbb{Z}_2$ symmetry corresponding to flipping all spins, and therefore a degenerate ground state.

To break this symmetry, the last qubit can be fixed as having spin $+1$, transforming the system into one with $n-1$ qubits where $h_i = J_{n-1,i}$. In effect, this allows SK problems to be created by sampling both $h_i$ and $J_{i,j}$ from $N(0,1)$. The inverse transform is also possible; an Ising Hamiltonian can be written as a system with $n + 1$ qubits, where $J_{n,i} = h_i$ and the new system has $h_i = 0$.
\begin{figure}[ht]
    \centering
    \[
    \underbrace{
    J =\begin{pmatrix}
    0      & J_{1,0}      & \cdots & J_{n-1,0} \\
    J_{1,0} & 0      & \cdots & J_{n-1,1} \\
    \vdots & \vdots & \ddots & \vdots \\
    J_{n-1,0} & J_{n-1,1} & \cdots & 0
    \end{pmatrix}, 
    \mathbf{h} = 
    \begin{pmatrix}
    h_0 \\
    h_1 \\
    \vdots \\
    h_{n-1}
    \end{pmatrix}
    }_{\text{Ising with Fields ($n$)}}
    \quad \iff \quad  J' =
    \underbrace{
    \left(
    \begin{array}{c|cccc}
    0      & h_0      & h_1      & \cdots & h_{n-1}    \\ \hline
    h_0    & 0      & J_{1,0}      & \cdots & J_{n-1,0}    \\
    h_1    & J_{1,0} & 0      & \cdots & J_{n-1,1}    \\
    \vdots & \vdots & \vdots & \ddots & \vdots \\
    h_{n-1}    & J_{n-1,0} & J_{n-1,1} & \cdots & 0
    \end{array}
    \right),
    \mathbf{h'} = \mathbf{0}
    }_{\text{Homogenized Ising ($n+1$)}}
    \]
    \caption{Mapping of an Ising model with external fields to and from a system with no external fields. The energy levels are the same in both Hamiltonians, but doubly degenerate in the homogenized system.}
    \label{fig:ising_homogenization}
\end{figure}
This relationship is made clearer in Figure \ref{fig:ising_homogenization}, where $J$ is represented by a symmetric matrix with $0$ on the diagonal.
This means the $n$-qubit Ising Hamiltonian can also be written as 
\begin{equation}
\label{Ising}
    \hat{H}_I = - \frac{1}{2}\sum_{i = 0}^{n-1} \sum_{j = 0}^{n-1} J'_{i,j} \hat{Z}_i \hat{Z}_j,
\end{equation}
where the sum is now over all $n^2$ elements.
This form will be useful for statistical analysis later.

There are two datasets used in this work. $\mathcal{D}_{\text{random}}$ is a dataset of SK problems for $n$ up to 20, and has been produced as part of \cite{data}. $\mathcal{D}_{\text{hard}}$ is a dataset of SK problems with a small minimum energy gap, which has been curated in \cite{TimData}.
The code developed for this paper can be found at \cite{MultistageQW}.

Following the conventions in \cite{Chancellor2022}, the parameters $J_{i,j}$ and $h_i$ will be treated as dimensionless, making $\hat{H}_I$ dimensionless too. All practically applied Hamiltonians will take the form 
\begin{align}
\label{Hamiltonian}
    \hat{H}(t) = B(t)\hat{H}_I-A(t) \hat{H}_G,
\end{align}
where $\hat{H}_G$ is a driver Hamiltonian, and the schedules $A(t)$ and $B(t)$ provide the energy scale. $A(t)$ and $B(t)$ are hardware dependent, and here a theoretical machine is considered with the same field strengths available as a D-Wave Advantage2 1.6 machine \cite{schedule}, but with no other restrictions on schedule shape or anneal times.

\section{Methods} \label{sec:methods}

We first precisely define what a multi-stage quantum walk is, then derive a method of choosing the free parameters that MSQW requires, and finally describe the simulation techniques used to study it. Our method is chosen to only require a polynomial amount of classical effort to choose parameters, since many methods require an exponential number of runs of a polynomial time algorithm in order to tune parameters \cite{Dennis} \cite{QAOA} \cite{VQE}.

\subsection{The Multi-Stage Quantum Walk}

Let $\hat{H}_I$ be a given problem encoded into an Ising Hamiltonian using $n$ qubits with $N = 2^n$ states, and then define the driver Hamiltonian as 
$$\hat{H}_G = \sum_{j=0}^{n-1} \hat{X}_j,$$
where the action of $\hat{X}_j$ is to flip qubit $j$. This is equivalent to the adjacency matrix of the $n$-dimensional hypercube graph, so will be referred to as the graph Hamiltonian. A quantum walk is then defined as the evolution of an initial state $\ket{\psi_0}$ by a time-independent Hamiltonian of the form in Equation \eqref{Hamiltonian}. This can be parameterised by a ratio $\gamma = \frac{B}{A}$ and an evolution time $t$ to give
\begin{equation} \label{walk}
\ket{\psi_1} = \exp \left( {-itA(0)\hbar(\hat{H}_I - \gamma\hat{H}_G)} \right)\ket{\psi_0}.
\end{equation} 
For simplicity, units where $A(0) \hbar = 1$ will be used from now. For the multi-stage case with $m$ stages, each stage $k$ of the quantum walk needs a pair of parameters $\gamma_k$ and $t_k$, resulting in
\begin{align}
    \label{QW}
    \ket{\psi_m} &= e^{-it_m\hat{H}_m} \ket{\psi_{m-1}} \\
    &= e^{-it_m\hat{H}_m} \cdots e^{-it_1\hat{H}_1}\ket{\psi_0},
\end{align}
where $\hat{H}_i = \hat{H}_I - \gamma_i\hat{H}_G$. 
In this work, $\ket{\psi_0}$ is chosen to be the uniform superposition, as this is the lowest energy state of $-\hat{H}_G$.
After the the last stage is complete, the state $\ket{\psi_m}$ is measured. Let $\ket{E_0}$ be the ground state of $\hat{H}_I$, then the probability that $\ket{\psi_m}$ is found to be in the ground state is $|\braket{E_0|\psi_m}|^2$.

\subsection{Heuristic {$\gamma_k$} values}
The definition in Equation \eqref{QW} is usable, but requires choosing values of $t_k$ and $\gamma_k$. To motivate how to choose these parameters, the infinite time average probability introduced in \cite{Callison} is extended to an arbitrary number of stages. The details are in Appendix \ref{InfTime}, with the main result being that 
\begin{align}
P_{\infty} &= \lim_{t \rightarrow \infty} \frac{1}{t^m} \int_0^t \cdots \int_0^t |\braket{E_0|\psi_m}|^2 dt_1 \cdots dt_m \\
&= \sum_{a_1 \cdots a_m = 0}^N  \left| \braket{E_0|E^{(m)}_{a_m}} \braket{E^{(m)}_{a_m} | E^{(m-1)}_{a_{m-1}}} \cdots \braket{E^{(1)}_{a_1}|\psi_0} \right|^2, 
\label{P_inf}
\end{align}
where superscripts are being used to denote that $\ket{E_{a_k}^{(k)}}$ are the $N$ eigenvectors of $\hat{H}_k$.

In this context, Equation (\ref{P_inf}) looks like a series of projections, which gradually rotate the initial state to the final state. To maximise the product of these terms, each projection should rotate the state the same amount towards $\ket{E_0}$. This can be achieved if $\hat{H}_k$ could be constructed such that
\begin{equation}
\label{spectralH}
    \sin \left(\frac{k \pi}{2(m+1)}\right) \ket{E_0} - \cos\left(\frac{k \pi}{2(m+1)}\right) \ket{\psi_0}
\end{equation}
is an eigenvector of it. It is possible to choose any relative phase factor between the terms in Equation \eqref{spectralH}; a phase of $\pi$ is chosen to match Equation \eqref{Hamiltonian}.

This can be done exactly in the case of $2 \times 2$ Hermitian matrices, see Appendix \ref{Example}. The formula that results is not applicable to larger problems, but it makes it clear that first it is necessary to scale the operators such that they have the same energy spread. The heuristic used here is
\begin{equation}
\label{spectral}
    \hat{H}_k = \sin\left(\frac{k \pi}{2(m+1)}\right) \Tilde{H}_I - \cos\left(\frac{k \pi}{2(m+1)}\right) \Tilde{H}_G,
\end{equation}
where the tilde denotes normalisation of energy spreads. This heuristic is spherical linear interpolation between $\hat{H}_I$ and $\hat{H}_G$, which tries to mimic the gradual rotation in Equation \eqref{spectralH}. For $\hat{H}_G$, its energy spread is simply $n$, but to normalise $\hat{H}_I$ exactly it would be necessary to solve the problem it encodes to calculate the energy spread. Similarly, methods such as simulated annealing can approximate the energy spread by finding an approximate solution.
The method used here is described in Appendix \ref{Stats}, which uses $J$ to calculate the variance of the energy distribution, and then fits a normal distribution from which the expected minimum energy is calculated. In principle, higher moments can be used, but in practice, as quantum walks are not sensitive to small deviations from optimal parameters \cite{Callison} \cite{Lasse}, there is no need for a more accurate normalisation.

Putting all this together, and scaling so that the coefficient of $\hat{H}_I$ is 1 gives
\begin{align} \label{gamma}
    \hat{H}_k = \hat{H}_I - \underbrace{\frac{\braket{E_{N-1} - E_0}}{2n}\cot\left(\frac{k \pi}{2(m+1)}\right)}_{\color{red}{\gamma_k}} \hat{H}_G.
\end{align}
A similar method is used in \cite{Callison} where the expected variance of a SK spin glass problem is used to estimate the energy spread, with a scaling factor to account for the non-normal tails.

\subsection{Heuristic {$t_i$} Values} \label{Timescales}
The previous subsection derived heuristic values for the hopping rates $\gamma_i$ in the infinite time limit.
To run the algorithm on a real quantum machine will clearly require choosing finite values for $t_i$ instead. In \cite{Callison}, the method used is to successively double $t$ until the success probability changes less than 5\% from the previous value and then analyse the results to find how $t$ scales with $n$. Here an analytic approach is taken to find the shortest viable evolution time.

Consider a single stage quantum walk, evolved under $\hat{H} = \hat{H}_I - \gamma \hat{H}_G$. \citeauthor{Perspectives} \cite{Perspectives} note that, since $\ket{\psi_0}$ is the ground state of $-\hat{H}_G$, time evolution must increase the energy with respect to the driver Hamiltonian $-\gamma\hat{H}_G$, and hence reduce it with respect to the problem Hamiltonian $\hat{H}_I$ due to conservation of energy.  We use the graph energy $\bra{\psi(t)} \hat{H}_G \ket{\psi(t)}$ as our metric, which begins at a maximum due to dropping the minus sign in the full Hamiltonian.  As $t$ increases, the graph energy must therefore decrease initially. Explicitly, the energy at time $t$ can be written as
\begin{align*}
E_G &= \bra{\psi_0} \exp(i\hat{H}t) \hat{H}_G \exp(-i\hat{H}t) \ket{\psi_0} \\
   &= \bra{\psi_0} (1 + i\hat{H}t - \frac{1}{2}\hat{H}^2t^2) \hat{H}_G (1 - i\hat{H}t - \frac{1}{2}\hat{H}^2t^2) \ket{\psi_0} +O(t^4) \\
   &= \bra{\psi_0} (\hat{H}_G - i\hat{H}_G\hat{H}t - \frac{1}{2} \hat{H}_G \hat{H}^2 t^2 + i\hat{H} \hat{H}_G t + \hat{H} \hat{H}_G \hat{H} t^2 - \frac{1}{2} \hat{H}^2 \hat{H}_G t^2) \ket{\psi_0} + O(t^4).
\end{align*}
By splitting this term into multiple inner products, it is then possible to reverse the order of terms in a given inner product by taking the conjugate transpose. Since each term represents an observable, they must be real so this doesn't change the value, and since the matrices are Hermitian they are also unchanged. This lets some terms cancel, giving
\begin{align*}
E_G =&\, \bra{\psi_0} \hat{H}_G \ket{\psi_0} - \bra{\psi_0} \hat{H}_G \hat{H}^2 t^2 \ket{\psi_0} + \bra{\psi_0} \hat{H} \hat{H}_G \hat{H} t^2 \ket{\psi_0} + O(t^4). \\
=&\, n - nt^2\bra{\psi_0} \hat{H}^2 \ket{\psi_0} + t^2\bra{\psi_0} \hat{H} \hat{H}_G \hat{H} \ket{\psi_0} + O(t^4). \\
=&\, n - nt^2 \bra{\psi_0} \gamma^2 \hat{H}_G^2 - \gamma \hat{H}_I\hat{H}_G - \gamma \hat{H}_G \hat{H}_I \\
&+ \hat{H}_I^2 \ket{\psi_0} + t^2\bra{\psi_0} \gamma^2 \hat{H}_G^3 - \gamma \hat{H}_G^2 \hat{H}_I - \gamma \hat{H}_I \hat{H}_G^2 + \hat{H}_I \hat{H}_G \hat{H}_I \ket{\psi_0} + O(t^4)\\
=&\, n - \gamma^2 n^3 t^2 - 2 \gamma n^2t^2 \bra{\psi_0} \hat{H}_I \ket{\psi_0} - nt^2 \bra{\psi_0} \hat{H}_I^2 \ket{\psi_0} + \gamma^2 n^3t^2 + 2 \gamma n^2t^2 \bra{\psi_0} \hat{H}_I \ket{\psi_0} \\
&+ t^2 \bra{\psi_0} \hat{H}_I \hat{H}_G \hat{H}_I \ket{\psi_0} + O(t^4)\\
=&\, n - nt^2 \bra{\psi_0} \hat{H}_I^2 \ket{\psi_0} + t^2 \bra{\psi_0} \hat{H}_I \hat{H}_G \hat{H}_I \ket{\psi_0} +O(t^4) .
\end{align*}

Only energies in $[-n,n]$ are physically viable, so when the prediction falls outside this range it can be assumed the energy has saturated. This happens when $E_G = -n,$ giving
\begin{align}
\label{time}
    t_s = \sqrt{\frac{2n}{n \bra{\psi_0} \hat{H}_I^2 \ket{\psi_0}-\bra{\psi_0} \hat{H}_I \hat{H}_G \hat{H}_I \ket{\psi_0}} }.
\end{align} 
Since $\ket{\psi_0}$ is the uniform superposition and $\hat{H}_I$ is diagonal, the first term in the denominator of Equation \eqref{time} is just the average of the squared energy levels. The action of $\hat{H}_G$ on a vector is to add together all elements with an index one bit flip away from each entry, so each entry of $\hat{H}_G \hat{H}_I \ket{\psi_0}$ will contain the sum of the $n$ energy levels adjacent to it. The entries of $\hat{H}_I \hat{H}_G \hat{H}_I \ket{\psi_0}$ are then a correlation between the energy levels and their neighbours. This allows the second term in the denominator of Equation \eqref{time} to be calculated. Let $\Delta_{k,l} = E_l - E_k$ if $E_l$ and $E_k$ are adjacent, and $0$ otherwise. Since $\Delta_{k,l} = -\Delta_{l,k}$, the sum can be written by only iterating over the terms with positive $\Delta_{k,l}$. This gives
\begin{align}
\nonumber \bra{\psi_0} \hat{H}_I \hat{H}_G \hat{H}_I \ket{\psi_0} &= \frac{1}{N}\sum_{k,l \text{ s.t } \Delta_{k,l} > 0} E_k (E_k + \Delta_{k,l}) + E_l (E_l - \Delta_{k,l}) \\
\nonumber &= \frac{1}{N}\sum_{k,l \text{ s.t } \Delta_{k,l} > 0} E_k^2 + E_l^2 +  (E_k - E_l) \Delta_{k,l}\\
\nonumber &= \frac{1}{N} \sum_{k,l \text{ s.t } \Delta_{k,l} > 0} E_k^2 + E_l^2  - \Delta_{k,l}^2 \\
\label{corr}
&= -\frac{1}{2} \braket{\Delta_{k,l}^2} + \frac{1}{N} \sum_{k,l \text{ s.t } \Delta_{k,l} > 0} E_k^2 + E_l^2 .
\end{align}
Since each energy level appears $n$ times in Equation \eqref{corr}, this part cancels the $\hat{H}_I^2$ term in Equation \eqref{time}, giving
\begin{align}
\label{time2}
    t_s = \sqrt{\frac{4n}{\braket{\Delta_{k,l}^2}}},
\end{align}
For a given $J$ matrix and $h$ vector of the form in Equation \eqref{IsingJ}, it is shown in Appendix \ref{Stats} that
\begin{align}
\label{delta}
\braket{\Delta_{k,l}^2} = 8 \sum_{i,j} J_{i,j} J_{i,j} + 4\sum_i h_j^2.
\end{align}
For SK spin glasses, the expected value of $\braket{\Delta_{k,l}^2}$ can be calculated using a result from \cite{Perspectives}. In that paper, it is calculated that $\Delta_{k,l}$ is normally distributed with mean 0 and standard deviation $\sqrt{2(n + 1)}$. The sum of all $Nn$ squared energy differences is, by definition, a scaled chi-squared distribution with $Nn$ degrees of freedom. As such, the expected value of $\braket{\Delta_{k,l}^2}$ is $2n(n+1)$, giving
\begin{align} \label{avgtime}
\braket{t_s} &= \sqrt{\frac{2}{(n+1)}}.
\end{align}
This same result can also be calculated from Equation \eqref{delta}.

This analysis only holds for the first stage of the quantum walk, however, it is also possible to derive a similar expression for the final stage. By starting in state $\ket{E_0}$ and expanding $\braket{\psi(t) |\hat{H}_I|\psi(t)}$ in the same way, it can be shown that
\begin{align}
\label{E_I}
    E_I = E_0 + \gamma^2 t^2 \left( \braket{E_0 | \hat{H}_G \hat{H}_I \hat{H}_G |E_0} - E_0 \braket{E_0 | \hat{H}^2_G |E_0} \right).
\end{align}
It is convenient to write this in terms of $E_G$. Due to the conservation of energy of $\hat{H}$, $\frac{dE_I}{dt} = -\gamma\frac{dE_G}{dt}$, a change in $E_I$ of $2n \gamma$ will cause a change in $|E_G|$ of $2n$, the maximum possible. Therefore,
$$2n\gamma = \gamma^2 t_f^2 \left( \braket{E_0 | \hat{H}_G \hat{H}_I \hat{H}_G |E_0} - E_0 \braket{E_0 | \hat{H}^2_G |E_0} \right). $$

The other terms can now be simplified. Since $\hat{H}_G$ is an adjacency matrix, the entries of $\hat{H}^2_G$ describe how many paths of length 2 exist connecting each pair of states, and in particular each state has $n$ ways of connecting to itself with two bit flips. This means $\braket{E_0 | \hat{H}^2_G |E_0}= n$. Finally, $\hat{H}_G \ket{E_0}$ is a vector with a 1 in the entries that are adjacent to the ground state, so $\braket{E_0 | \hat{H}_G \hat{H}_I \hat{H}_G |E_0}$ is the sum of energies of the states adjacent to the ground state. Writing each of these states as $E_0 + \Delta_{0,k}$, then the final stage time heuristic can be written as
\begin{align}
    t_f = \sqrt{\frac{2n}{\gamma \sum_l \Delta_{0,l}}}.
\end{align} 

For a Hamiltonian of the form in Equation \ref{Ising}, then it can be seen in Equation \eqref{energydiff} that flipping spin $a$ of a classical state $\vec{s}$ leads to an energy change of $-2\vec{s}_a (J\vec{s})_a$. The sum over all spin flips is then $-2 \vec{s}^T J \vec{s} = 4E$ where $E$ is the state energy. This can be estimated with the same energy spread heuristic as used in Appendix \ref{heuristic}, since $E_0$ will be approximately half the energy spread, but choosing an evolution time that is too short reduces success probabilities drastically, whereas choosing an evolution time that is too long has no such effect. As such, it is important to always overestimate the evolution time, so one quarter of the energy spread will be used to estimate $E_0$. This gives
\begin{align*}
    \sum_l \Delta_{0,l} \approx \braket{E_{N-1} - E_0}.
\end{align*}
The final stage time heuristic is then
\begin{align}
\label{time3}
    t_f = \sqrt{\frac{2n}{\gamma \braket{E_{N-1} - E_0}}}.
\end{align}

Both $t_s$ and $t_f$ become overestimates for a large number of stages, since it's unrealistic to expect every stage to cause the maximum possible change in $E_G$. Instead, the expected change in $E_G$ is estimated from the same assumption of equal rotation between each stage used in Equation \eqref{spectralH}. Specifically, 
\begin{align}
    \Delta E_i = 2n \left( \frac{\gamma_{i-1}}{\sqrt{1 + \gamma^2_{i-1}}} - \frac{\gamma_{i+1}}{\sqrt{1 + \gamma^2_{i+1}}} \right),
\end{align}
where $\gamma_0$ is considered to be sufficiently large such that $\sqrt{1 + \gamma^2_0} \approx \gamma_0$ and $\gamma_{m+1} = 0$ so the single stage case still predicts the maximum change of $2n$. The heuristic used for every stage is then
\begin{align}
\label{time_final}
    t_i = \max \left( \sqrt{\frac{2\Delta E_i}{\braket{\Delta_{k,l}^2}}}, \sqrt{\frac{\Delta E_i}{\gamma_i \braket{E_{N-1} - E_0}}}\right).
\end{align}

Similar formulae appear in \cite{Dennis} where the Rabi oscillations caused by local Hamiltonians are considered, expanding Equation (14) of that paper to lowest order would give Equation \eqref{E_I} if the initial state were $\ket{k}$. In light of this, the two heuristics derived here can be viewed as the time needed for $\gamma \hat{H}_G$ to drive any eigenstate of $\hat{H}_I$ to an adjacent eigenstate, and the time needed for $\hat{H}_I$ to drive any eigenstate of $\gamma \hat{H}_G$ to an adjacent eigenstate. The overall heuristic in Equation \eqref{time_final} allows enough time for the slower of these two to occur. This also indicates that $O(n)$ stages are needed, since each stage allows time for a single bit flip to occur, and $n$ bit flips allows every state to be reached from any starting state. It may be that $\frac{n}{2}$ stages is closer to optimal in terms of time-to-solution, as it allows half of all states to be visited from a given starting state, and the usual equal superposition starting state has all qubits in a ``half flipped'' state.

\begin{figure}
\includegraphics[width=0.49\linewidth]{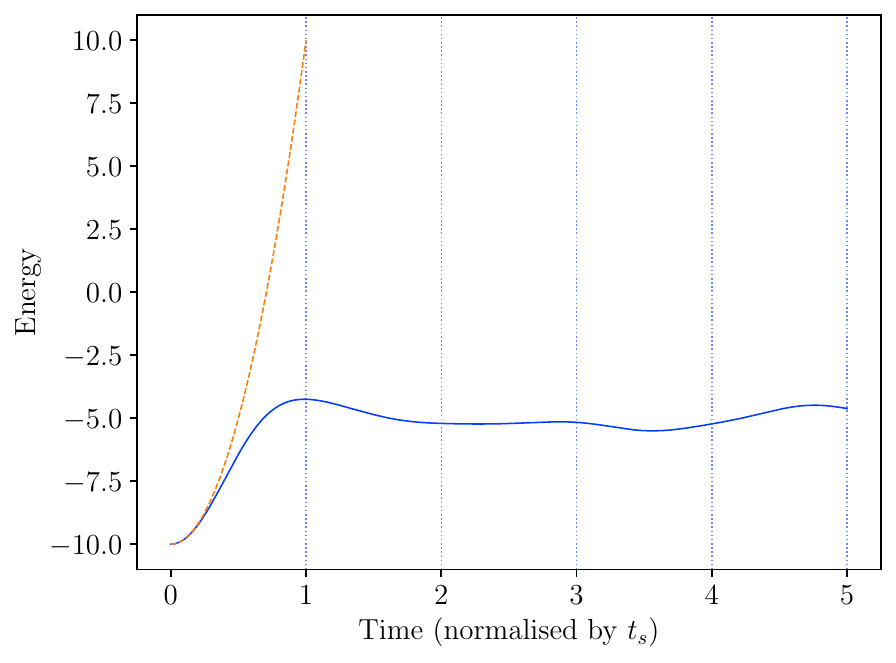}
\fcaption{The evolution of $\ket{\psi_0}$ for the 10 qubit problem `aaaufeflfwqdwhthcrcnynopihzciv', showing how well $t_s$ (dotted blue line) predicts the end of the fast growth period for the energy of $\hat{H}_G$ (solid blue line). The 2nd order approximation to $E_G$ is shown with a dashed orange line.}
\label{Energy}
\end{figure}

\subsection{Numerical Methods}

In \cite{data}, a set of 10,000 SK spin glass problems have been created for each $n$ from 5 to 20. In this work, the first 2,000 entries for each $n$ from 5 to 18 have been used. The format of the dataset is a set of binary files containing a vector of length $n$ for $h$, and a matrix of dimensions $n \times n$ for $J$, however, the matrices stored are not symmetric. Let the matrix in the file be $A$, then the lower triangle of $\frac{A + A^{T}}{2}$ is the $J$ used in Equation \eqref{IsingJ} in order to match the results in \cite{Callison} and \cite{Perspectives}. Since the entries of $A$ were drawn from a standard normal distribution, the entries of $J$ follow $N(0,\frac{1}{2})$ rather than a standard normal. This can be fixed by simply scaling up $J$ by $\sqrt{2}$, but it is argued in \cite{Callison} that this doesn't jeopardise the difficulty of the problems.

A second dataset, containing SK spin glass problems that have been curated to have a small minimum energy gap, has been provided in \cite{TimData}. The format of this dataset is a set of .h5 files which contain the $J$ and $h$ values as data fields. It could be argued that the post-selection of problems to have a small energy gap means they are no longer SK spin glass problems, but that is not important here.

For $n \leq 10$, a technique described in Appendix \ref{InfTime} has been used to calculate the infinite time average in Equation \eqref{P_inf} by diagonalising each $\hat{H}_i$. For $n \leq 18$, the heuristic time $t_i$ is calculated from Equation \eqref{time_final}, and then used to calculate the short time average success probability 
\begin{align}
P(t_1, t_2, ..., t_m) &= \left( \prod_{i=1}^m{t_i} \right)^{-1} \int_{t_m}^{2t_m} \cdots \int_{t_1}^{2t_1} \braket{E_0|\psi_m}\braket{E_0|\psi_m}^* dt_1 \cdots dt_m.
\end{align}
This integral is estimated via a Monte-Carlo method, where the time for each stage is randomly sampled in $[t_i,2t_i]$, and then repeated 100 times to give a reasonable statistical error.

The finite time evolutions were calculated by using the Carathéodory-Fejér method \cite{CF} to create a polynomial $p$ which approximates $\exp(-ix)$ to a relative error of $10^{-6}$ for all $x$ within the spectral radius of $\hat{H}$, $\rho(\hat{H})$. The number of terms needed is proportional to $\rho(\hat{H})$, which grows like $O(\sqrt{n^3 \ln(n)})$ according to the heuristic in Appendix \ref{heuristic} when applied to spin glasses. For the values of $n$ used in this work, polynomials of degree up to a few hundred are needed. As such, numerical accuracy is limited by the use of single precision floating point, which provides relative precision of roughly $5 \times 10^{-8}$ \cite{float}, with errors building up with successive operations. This is deemed sufficient, as the Monte-Carlo sampling calculates success probabilities with an expected relative error of roughly $4\%$, much larger than any numerical errors even in single precision.

This is then used to evaluate $\exp(-i\hat{H})\ket{\psi_0}$ as $p(\hat{H})\ket{\psi_0}$. It is possible to evaluate this efficiently using only matrix-vector multiplications by applying a modified version of the Clenshaw algorithm \cite{Clenshaw}, showing better convergence and numerical stability than a Taylor series or numerical integration approach. More details and a formal error analysis will be given in an upcoming paper \cite{Numerics}. For Figure \ref{anneal}, a quantum anneal has been simulated using QuTiP \cite{QuTiP}, and for Figure \ref{anneal_median} a numerical method for breaking quantum anneals into quantum walks has been used \cite{CFET}.

\section{Results and Discussion} \label{sec:results}

\begin{figure}
\includegraphics[width=0.49\linewidth]{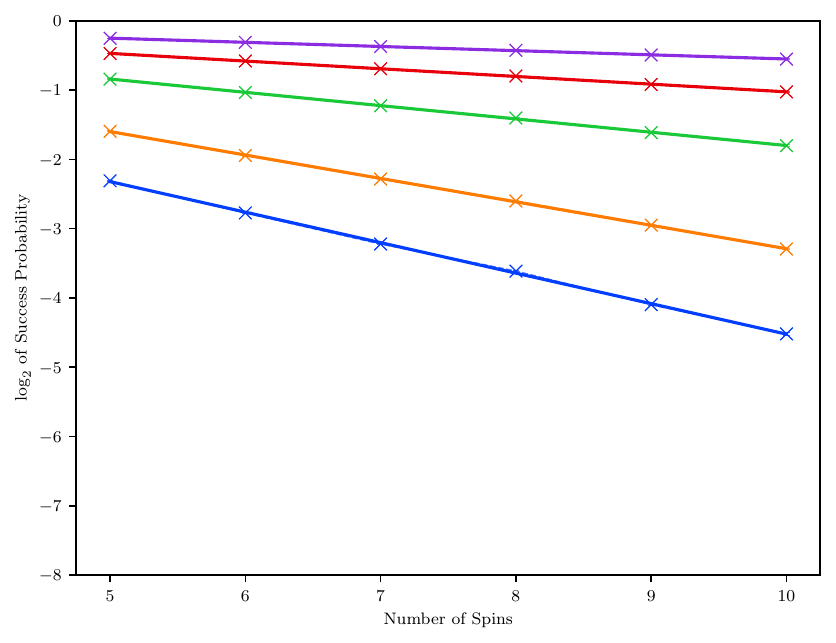}
\fcaption{The median success probability in the infinite time average for $\mathcal{D}_{\text{random}}$. Each line represents a multi-stage quantum walk with a different number of stages. From bottom to top, the stages shown are 1 (blue), 2 (orange), 5 (green), 10 (red) and 20 (purple). Errors shown are standard errors on the median and have been calculated via bootstrap sampling with 1000 samples per point. Solid lines show regression lines, dotted lines guide the eye.}
\label{inf_median}
\end{figure}

\begin{figure}
\includegraphics[width=0.49\linewidth]{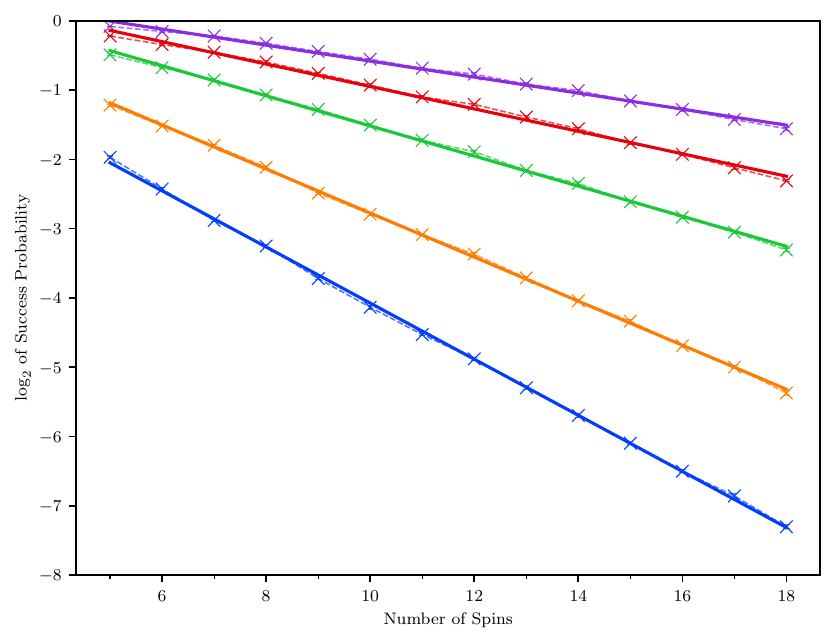}
\fcaption{The median success probability using the heuristic time from Eq.~\eqref{time_final} across a range of system sizes for $\mathcal{D}_{\text{random}}$. Each line represents a multi-stage quantum walk with a different number of stages. From bottom to top, the stages shown are 1 (blue), 2 (orange), 5 (green), 10 (red) and 20 (purple). Errors shown are standard errors on the median and have been calculated via bootstrap sampling with 1000 samples per point. Solid lines show regression lines, dotted lines guide the eye.}
\label{short_median}
\end{figure}

\begin{figure}
\includegraphics[width=0.49\linewidth]{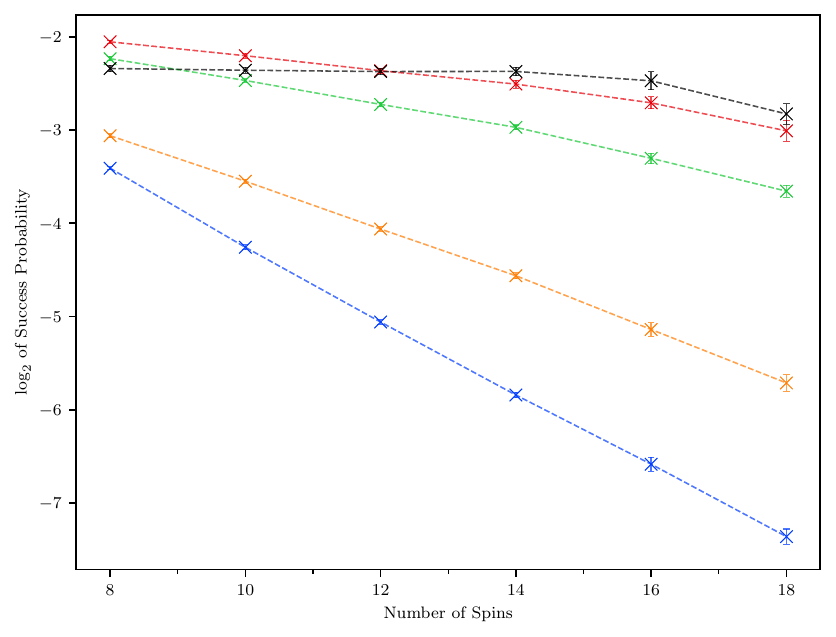}
\fcaption{The median success probability for $\mathcal{D}_{\text{hard}}$. Each line represents a multi-stage quantum walk with a different number of stages. From bottom to top at 16 spins, the stages shown are 1 (blue), 2 (orange), 5 (green), 10 (red) and 50 (black). 20 stages has been intentionally omitted for clarity. Errors shown are standard errors on the median and have been calculated via bootstrap sampling with 1000 samples per point. No regression lines have been calculated both due to poorness of fit and theoretical considerations mentioned in Section \ref{sec:results}.}
\label{hard_median}
\end{figure}

Three main sets of results have been calculated. The first calculates the infinite time probability from Appendix \ref{InfTime} for up to 20 stages on $\mathcal{D}_{\text{random}}$. This can be seen in Figure \ref{inf_median}. The other sets of results have been calculated using the heuristic time in Eq.~\eqref{time_final}. These are shown in Figure \ref{short_median} and Figure \ref{hard_median} for $\mathcal{D}_{\text{random}}$ and $\mathcal{D}_{\text{hard}}$ respectively. All points represent the median success probability for all problems in the respective dataset. The median has been used to better separate the behaviour of the method on typical and hard problems, as the behaviours are significantly different.

\begin{figure}
%\centering
\subfloat[\justifying The slope of the regression line in Equation \eqref{regress} with number of stages, indicating that adding more stages will improve the scaling of the method. From top to bottom at 10 stages, the orange line corresponds to the infinite time average on $\mathcal{D}_{\text{random}}$, and the blue line to the short time average on $\mathcal{D}_{\text{random}}$. Errors shown are the standard errors provided by \mono{scipy.stats.linregress}.]
{\includegraphics[width=0.49\linewidth]{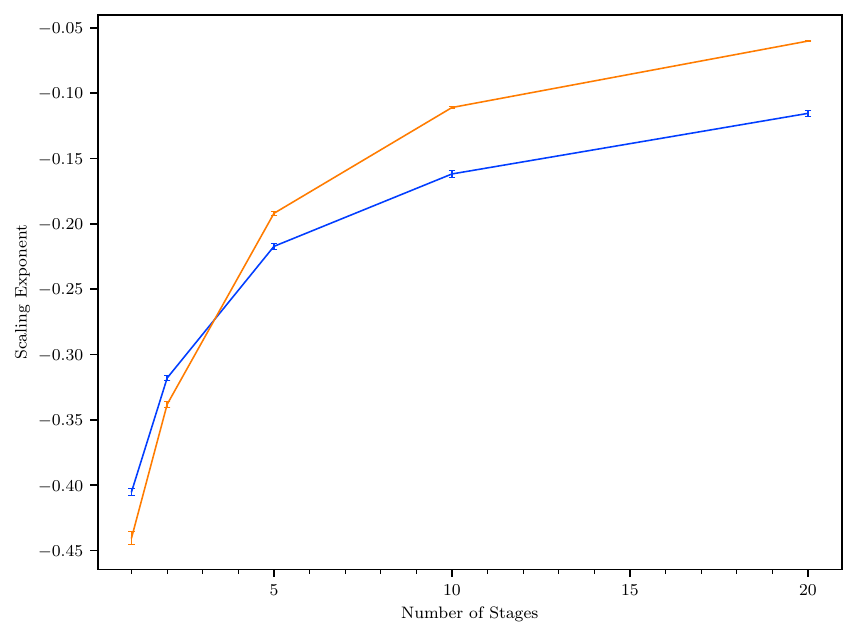}\label{scaling}}
\hfill
\subfloat[\justifying  The x-intercept of the regression line in Equation \eqref{regress}, indicating that the short time heuristic outperforms infinite time for small problems. From top to bottom, the blue line corresponds to the short time average on $\mathcal{D}_{\text{random}}$, and the orange line to the infinite-time average on $\mathcal{D}_{\text{random}}$. Errors shown are the standard errors provided by \mono{scipy.stats.linregress}]
{\includegraphics[width=0.49\linewidth]{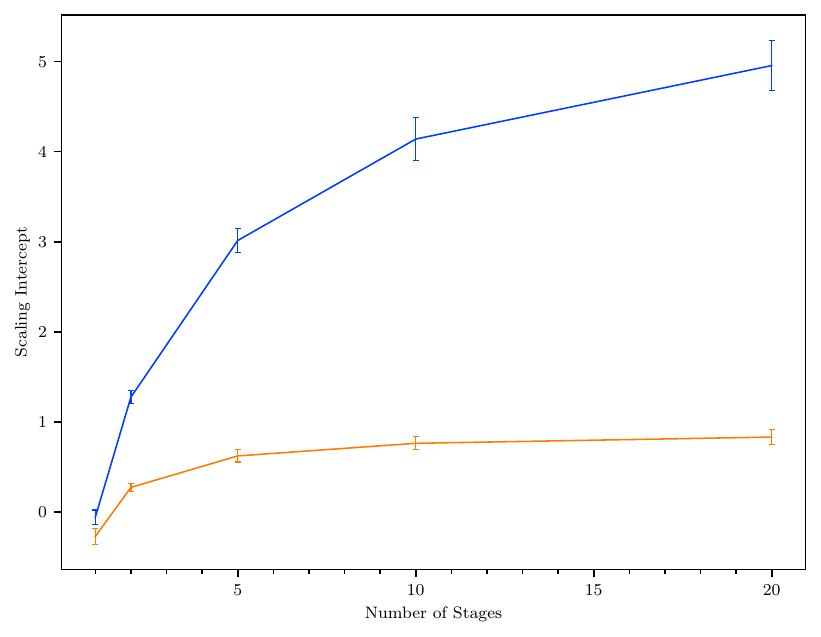}\label{scaling2}}
\fcaption{Results of the fitting of Equation \eqref{regress} to the data in Figure \ref{inf_median} and Figure \ref{short_median}.}
\label{fig:combined}
\end{figure}

The best known exact classical solver for SK spin-glass problems has a runtime of $O(2^{0.451n})$ in the average case \cite{bestquantum}. As such, any new algorithms are also expected to scale like $O(2^{an})$, where $a$ is algorithm dependent.
For quantum algorithms with probabilistic outcomes, it is more useful to discuss success probability, given by $O(2^{-an})$, with $O(2^{an})$ runs needed on average to find a solution. Each individual run of the algorithm will have a runtime that can scale with $n$, with the exact scaling often being machine dependent. For the algorithms being discussed here, that scaling is a small polynomial. The best known quantum algorithm not based on quantum walks applies quantum branch-and-bound to the best known classical algorithm, to give $a = 0.226$ \cite{bestquantum}. Quantum walks offer an alternative solution pathway, with previous work achieving $a = 0.410$ \cite{Callison} with a single stage quantum walk, and $a = 0.12$ with guided quantum walks \cite{Dennis}.

For both the infinite time and short time cases, the success probability of the MSQW can be well fit by 
\begin{equation} \label{regress}
    P = \exp\left(-a(n - n_0)\right)
\end{equation}
for some functions $a$ and $n_0$ which depend on the number of stages $m$. $n_0$ captures the constant factor hidden by the big-O notation, but written in this way it can also be interpreted as the size of problem which is trivial in the average case. It can be seen in Figure \ref{scaling} that adding more stages decreases the scaling exponent both in the average case and the infinite time case, giving a faster algorithm overall. In Figure \ref{scaling2}, it can be seen that adding more stages increases $n_0$ for the short time heuristic too. This shows its ability of it to stop at a peak of the success probability, rather than integrating over both peaks and troughs as the infinite time average does.

At these problem sizes, for random problem instances, adding more stages increases success probability without limit, but this view is overly optimistic. When MSQW is tested on $\mathcal{D}_{\text{hard}}$, it can be seen that a limit is hit where adding more stages decreases success probability. This limit depends on both $n$ and the minimum gap of the problem. This is also the expected behaviour of the algorithm on typical problems as $n \rightarrow \infty$, since as $n$ increases, the proportion of problems with a small energy gap also increases. This is shown in Table \ref{tab:hard_rates}, which has been calculated with the code provided at \cite{MinGap}, and this is also noted in \cite{TimData}. This shows that an absolute value cut-off does not correctly capture instance hardness as size increases, and hardness should instead be measured via percentiles as has been done in \cite{Mirkarimi2023}.
\begin{table}[h]
\centering
\begin{tabular}{|c|c|}
\hline
\textbf{n} & \textbf{Hard Problems (\%)}\\ \hline
6  & 1.58 $\pm$ 0.125 \\ \hline
8  & 2.83 $\pm$ 0.166 \\ \hline
10 & 3.81 $\pm$ 0.191 \\ \hline
12 & 5.26 $\pm$ 0.223 \\ \hline
14 & 5.91 $\pm$ 0.236 \\ \hline
\end{tabular}
\caption{Proportion of problems out of 10,000 typical instances with a minimum gap below 0.01}
\label{tab:hard_rates}
\end{table}
For this reason, regression fits have not been calculated for $\mathcal{D}_{\text{hard}}$. The rate at which the problems become more difficult with $n$ is skewed by the post-selection step to be smaller than for $\mathcal{D}_{\text{random}}$, which gives poor fits which paradoxically imply an algorithm that scales better for hard problems than typical ones. 
 
It is nevertheless clear that the chosen heuristic values work well without the need for a classical optimization process to refine the parameters $\gamma$ and $t$, as the short time heuristic gives similar scaling to the infinite time case. For $m = 1$, the success probability for $\mathcal{D}_{\text{random}}$ scales like $O(2^{-(0.404 \pm 0.003)n})$ in the short time case for dataset 1, in good agreement with the value of $O(2^{-(0.410 \pm 0.002)n})$ reported in \cite{Callison}. For $m=20$, the success probability scales like $O(2^{-(0.115 \pm 0.002)n})$, beating the previous known bests from \cite{bestquantum} and \cite{Dennis} if this scaling holds to larger $n$.

\begin{figure}
\includegraphics[width=0.49\linewidth]{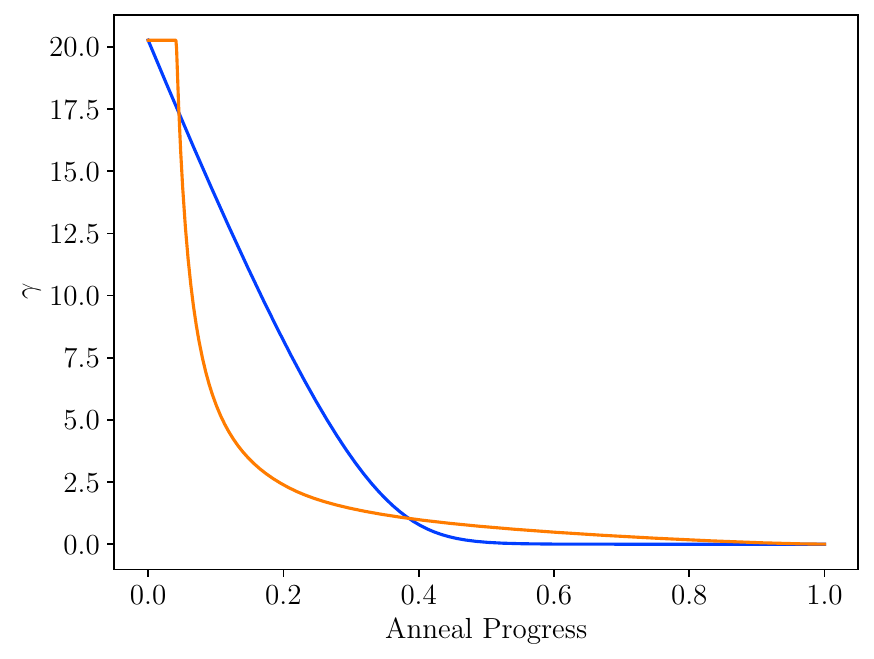}
\fcaption{A comparison of the D-Wave fast anneal schedule (blue) with the infinite-stage anneal schedule (orange). For the D-Wave schedule, $\gamma$ is effectively 0 at the halfway point, with half the evolution time used for freeze-out before measurement.}
\label{schedules}
\end{figure}

\begin{figure}
\begin{minipage}[c]{0.49\linewidth}
\includegraphics[width=\linewidth]{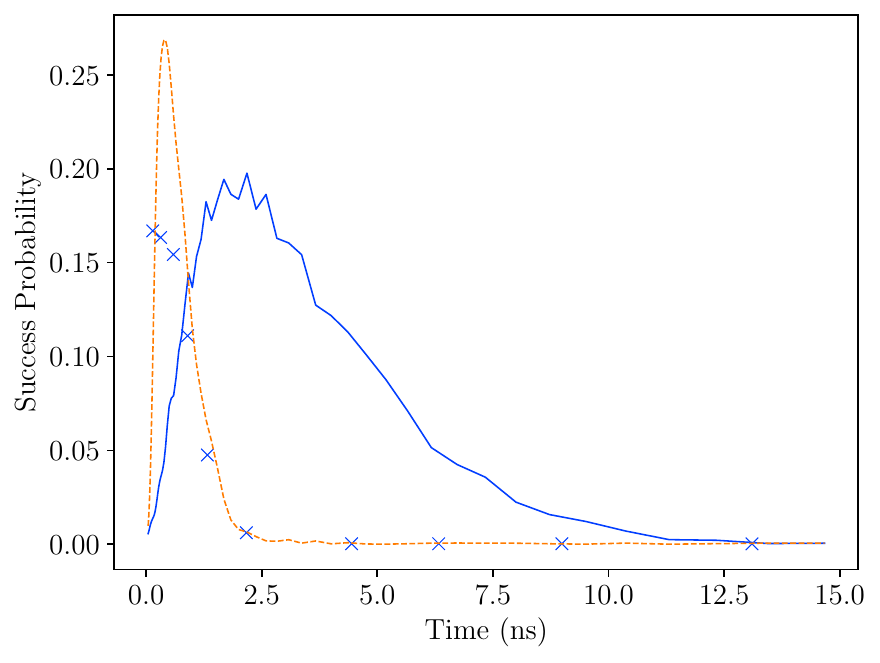}
\end{minipage}
\hfill
\begin{minipage}[c]{0.49\linewidth}
\includegraphics[width=\linewidth]{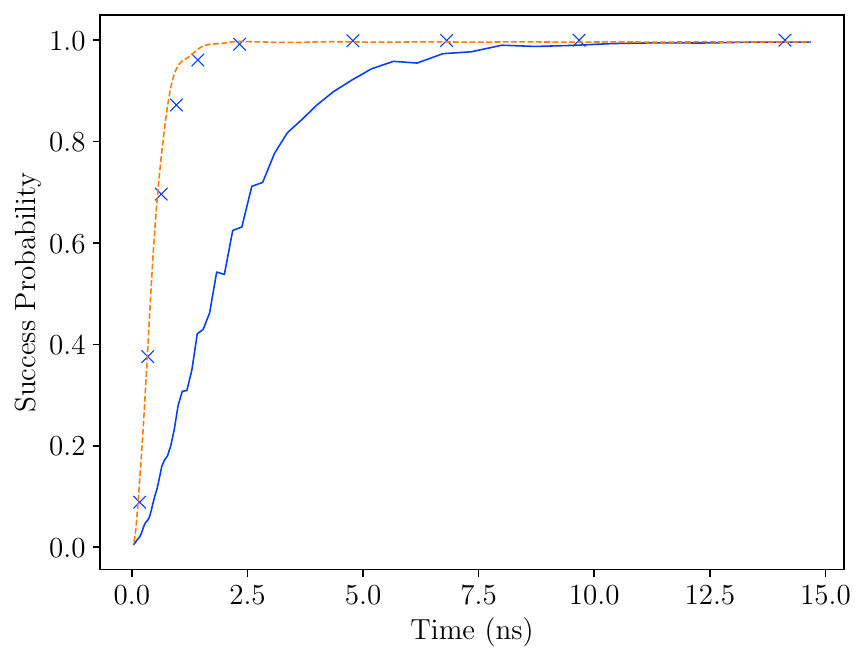}
\end{minipage}
\fcaption{The success probability for the 8-qubit problems 'epvftnkpdsdqbtsthxpsicmodcfnot' (left) and 'epxjlajlmdnmuzdjivgwgchldgisvv' (right) from \cite{data} when a quantum anneal is simulated using the same field strengths as a D-Wave Advantage2 1.6 machine for a range of total anneal times. The solid blue line is D-Wave's fast anneal schedule, whereas the dashed orange line is the infinite-stage schedule. MSQWs are indicated by crosses for 1, 2, 5, 10, 20, 50, 200, 400, 800 and 1600 stages. They achieve similar success probabilities in much shorter time. Although both problems are from the set of typical problems, the problem on the left has a very small minimum gap and is therefore a particularly difficult problem instance.}
\label{anneal}
\end{figure}

\begin{figure}
\includegraphics[width=0.49\linewidth]{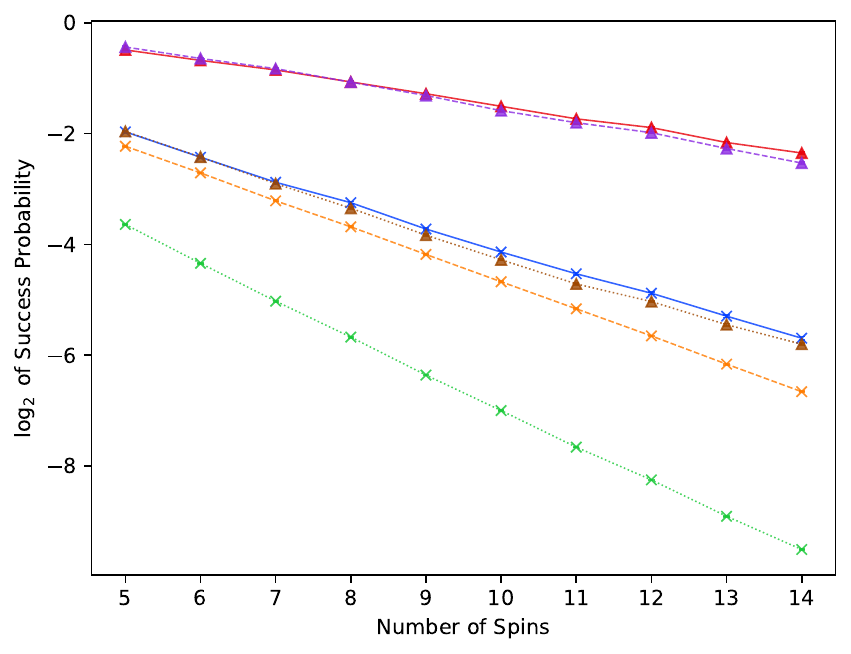}
\fcaption{Median success probability for MSQW compared against two smooth annealing schedules on $\mathcal{D}_{\text{random}}$, under two options for total time. Line style indicates method: quantum walks (solid), infinite-stage schedule (dashed), and D-Wave's fast anneal schedule (dotted). Marker shape indicates anneal time budget: crosses for equivalent time to the single-stage quantum walk, triangles for equivalent time to a 5-stage quantum walk. Notably, single-stage quantum walks outperform D-Wave's fast anneal schedule even when D-Wave is given a 5-stage anneal time budget.}
\label{anneal_median}
\end{figure}

Quantum walks approximate smooth anneal schedules when enough stages are used \cite{Lasse}, being essentially the tangent segments used in discrete approximation methods \cite{CFET}. As $m \rightarrow \infty$, the heuristics in Section \ref{sec:methods} become equivalent to an anneal schedule, which is shown in Figure \ref{schedules} for the problem instance shown in Fig \ref{anneal} (left). In these examples, 1600 stages have been used to estimate the infinite-stage schedule. This is excessive, for the problems analysed here the $4th$ order method from \cite{CFET} uses roughly 50 stages to match an anneal schedule. The analytics from \cite{Lasse} give an error of roughly $O(\frac{E_0^2}{m^2}) \approx O(\frac{n^3}{m^2})$ for MSQWs approximating an annealing schedule, but \cite{CFET} provides multiple alternative splitting schemes that achieve up to $8th$ order in accuracy. The $4th$ order method is used here as an error formula is provided, which allows rigorous bounding of errors without relying on asymptotic assumptions. Nevertheless, the complexity of calculating an $m$ stage schedule is $O(m)$, so using too many stages to calculate the infinite-stage schedule is not a concern.

The examples in \ref{anneal} are representative of the behaviour of the algorithm on problems with large and small minimum gaps. For problems with a small minimum gap, it is advantageous to stay in the diabatic regime as the time needed to reach the adiabatic regime is prohibitive \cite{Crosson2020}. The discrete nature of MSQWs can limit its coverage of the diabatic regime, in which case using the infinite-stage anneal schedule may be advantageous.

Both MSQWs and annealing with the infinite-stage schedule outperform D-Wave's annealing schedule significantly, as shown in Figure \ref{anneal_median}. An immediate advantage is the fact that the D-Wave schedule spends nearly half its time in the freeze-out regime, where the driver Hamiltonian is too small to produce significant change in the state, which is avoided by the heuristics used here. As expected, for a large number of stages the infinite-stage schedule becomes equivalent to MSQWs, but at a smaller number of stages MSQWs maintain better scaling.

There are a few obstacles to implementing MSQWs on a D-Wave machine. Firstly, D-Wave's fast anneal protocol requires $5\,ns$ minimum. Comparing with the timescales of Figure \ref{anneal} reveals that the ramping up of fields between walk stages cannot be ignored as they are not fast enough. A possible workaround for this is to reduce the $h$ and $J$ parameters in order to slow down the evolution, however this would reduce their precision. Nonetheless, this may be necessary to access the diabatic regime at all in some cases. Secondly, D-Wave's annealing controls only allow 10 points to be specified in the anneal curve, so the maximum number of walk stages possible is 5. This is another advantage of the infinite-stage schedule, as it can be more closely captured with just 10 points.

An alternative path to hardware implementation is to use Trotter decomposition to decompose each quantum walk stage into QAOA steps, which can then be implemented on a gate based quantum computer. This would avoid the obstacles above, but would likely require a large gate depth which is an obstacle of its own.

For the comparison in Figures \ref{anneal} and \ref{anneal_median}, it is necessary to convert the arbitrary units of $t_i$ into seconds. the field strengths $A(s)$ and $B(s)$ of a D-Wave Advantage2 1.6 machine have been used \cite{schedule}. Here $s$ is the anneal fraction, which starts at 0 and ends at 1. For the default D-Wave anneal, $s$ increases linearly from the start to the end of the anneal, but a new schedule can be described by specifying the function $s(t)$. $A(s)$ and $B(s)$ are provided in units of $\frac{\text{GHz}}{h}$, which are converted to $\frac{\text{GHz}}{\hbar}$ and substituted into Equation \eqref{walk}. D-Wave additionally restricts the $J$ and $h$ parameters to the intervals $[-4,4]$ and $[-2,1]$ respectively. It is therefore necessary to scale the parameters by some scale factor $\alpha$ to be within these bounds. For a random SK spin glass problem, $\alpha \sim  2\sqrt{\log(n)}$. This would cause $t_i$ to be multiplied by $\alpha$ on real hardware. Finally, for a given $\gamma_i$ it is necessary to look up the real value of the Ising field strength $A(s^*)$ at the point $s^*$ in the anneal schedule that achieves this ratio between the terms. This has been done here using cubic spline interpolation on the annealing schedule data in \cite{schedule}. Altogether this gives 
\begin{align}
t_{real} = \frac{\hbar \alpha}{A(s^*)}t_i.
\end{align}
This is then summed over all stages to calculate the times shown in Figure \ref{anneal}.

It is also worth noting that the problems described here have all-to-all connectivity, and real machines typically don't, so the actual Hamiltonian being implemented will not be the one being simulated here due to the need for minor embedding. This means an $n$-qubit all-to-all problem needs $O(n^2)$ physical qubits \cite{embedding} to implement, and while this should not affect the implementation in principle, this has not been investigated.

It can be seen from Equation \eqref{avgtime} that the running time of a single stage quantum walk for SK spin glasses is
$$T = O(n^{-0.5}),$$
much smaller than the mixing time in \cite{Callison} which found $T \propto n^{0.75}$ by direct measurement. That paper also shows that picking $T \propto n^{0.5}$ produces no measurable difference in success probability, so the shortest known time that achieves the infinite time average seems a factor of $n$ longer than the heuristic used here. This aligns with the interpretation of Equation \ref{time_final}, which indicates that $n$ stages would be needed for full mixing to be achieved, which would give a total running time of $O(n^{0.5})$. Our result is in line with the calculations of \cite{Banks}, which gives a bound of $O(n^{-1})$ on the minimum useful running time for a problem defined on a complete graph, and conjectures that the mixing time must then be at least $O(1)$.

It seems that using $t_i$ from Eq.~\eqref{time_final} produces worse scaling than the infinite time average for typical problems, as shown by Figure \ref{scaling}. Given the approach taken here, this is to be expected as Eq.~\eqref{time_final} is intended to be the shortest time that is still viable. As such, the expectation is the optimal evolution time in terms of time-to-solution is larger than or equal to the heuristic used here. For hard problem instances, however, it seems that shorter times are beneficial as longer evolution times seem to be more likely to end in the first excited state instead. For the example in Figure \ref{anneal} (right) it is necessary to go to an annealing time of $\sim 1$\ ms before the success probability starts to increase again in accordance with the adiabatic theorem.

\section{Conclusion and Future Work}\label{sec:further}

In this work, we have developed effective heuristics for choosing the free parameters of a MSQW with polynomial classical effort. We have shown numerically that these heuristics perform well for typical SK spin glass problems, giving a scaling exponent that improves polynomially with number of stages. This remains true for problem instances with a small minimum gap, however the scaling intercept decreases quickly enough to reduce success probability as the number of stages increases. We have also shown that MSQWs compare favourably to standard quantum annealing, achieving similar success probabilities in much less time. This shows that MSQWs with heuristically chosen parameters offer a practical approach to quantum optimization, and the tools we have presented here are general enough to be applied to a wide range of optimization problems encoded as Ising Hamiltonians.

Our numerical tests have necessarily been restricted to small problems, given the anneal times are too fast for currently available D-Wave hardware. We have therefore not been able to properly explore heuristics for determining the optimal number of stages. For a given finite sized problem, the optimal number of stages will be finite, enough to provide a close enough approximation to the optimal annealing schedule. As mentioned in Section \ref{Timescales}, there is reason to expect the optimal number of stages to be $\approx n$, however the precise number seems to be related to the energy gap, which cannot be known prior to solving the problem.  Although physical implementation on current D-Wave machines is unlikely, due to how short the needed timescales are, faster relative timescales may be available on Rydberg platforms. This would be an interesting direction for future work, to validate the theoretical and numerical results presented here.

There are a number of open questions around approximate vs exact solutions and the hardness of the same problem type under these different criteria \cite{zoufal}, including dependence on the type and tightness of the error bounds. It would be interesting to determine whether our heuristics work better for approximate solutions, as well as testing them on other problem types besides the Sherrington Kirkpatrick spin glasses.

\begin{acknowledgments}
We thank Tim Bode, Alexandru Ciobanu, Sebastian Schulz and Dongjin Suh for useful discussions and providing the data for the hard problem instances.
VK is supported by the EPSRC UK Quantum Technology Hubs in Quantum Computing and Simulation (EP/T001062/1) and Quantum Computing via Integrated and Interconnected Implementations (EP/Z53318X/1), and by UKRI EPSRC Collaborative Computational Project on Quantum Computing (EP/T026715/2).
AH was supported by the UKRI EPSRC International Network on Quantum Annealing (INQA) EP/W027003/1 for a visit to Forschungs Zentrum J\"ulich. AH is funded by UKRI EPSRC PhD studentship number 2745408.
\end{acknowledgments}

\bibliography{apssamp}

\appendix

\section{Infinite Time Probability} \label{InfTime}

This derivation follows the same logic as in \cite{Callison}. The first step is to write the operators in terms of their eigenvalues and eigenvectors. If there are $n$ qubits in the system, then the matrices have dimensions $N = 2^n$, giving
$$\hat{H}_k = \sum_{a_k=0}^{N-1} E^{(i)}_{a_k} \ket{E^{(k)}_{a_k}}\bra{E^{(k)}_{a_k}}.$$
This means
\begin{align*}
    \ket{\psi_m} &= \sum_{a_1 \cdots a_m = 0}^{N-1} e^{-it_mE^{(m)}_{a_m}} \cdots e^{-it_1E^{(1)}_{a_1}} \ket{E^{(m)}_{a_m}}\bra{E^{(m)}_{a_m}} \cdots  \ket{E^{(1)}_{a_1}} \bra{E^{(1)}_{a_1}}\ket{\psi_0},
\end{align*}
which can be rearranged as a series of inner products,
\begin{align*}
    \ket{\psi_m} &= \sum_{a_1 \cdots a_m = 0}^{N-1}  e^{-it_mE^{(m)}_{a_m}} \cdots e^{-it_1E^{(1)}_{a_1}} \braket{E^{(m)}_{a_m} | E^{(m-1)}_{a_{m-1}}} \cdots \braket{E^{(1)}_{a_1}|\psi_0} \ket{E^{(m)}_{a_m}}.
\end{align*}
The chance of measuring the ground state, $\ket{E_0}$, is then
\begin{align*}
 \braket{E_0|\psi_m}  \braket{E_0|\psi_m} ^* &= \left( \sum_{a_1 \cdots a_m = 0}^{N-1}  e^{-it_mE^{(m)}_{a_m}} \cdots e^{-it_1E^{(1)}_{a_1}} \braket{E^{(m)}_{a_m} | E^{(m-1)}_{a_{m-1}}} \cdots \braket{E^{(1)}_{a_1}|\psi_0} \braket{E_0 | E^{(m)}_{a_m}} \right) \\
 & \times \left( \sum_{b_1 \cdots b_m = 0}^{N-1}  e^{it_mE^{(m)}_{b_m}} \cdots e^{it_1E^{(1)}_{b_1}} \braket{E^{(m)}_{b_m} | E^{(m-1)}_{b_{m-1}}} \cdots \braket{E^{(1)}_{b_1}|\psi_0} \braket{E_0 | E^{(m)}_{b_m}} \right).
 \end{align*}
The infinite time average can then be calculated as 
$$ \lim_{t \rightarrow \infty} \frac{1}{t^m} \int_0^t \cdots \int_0^t \braket{E_0|\psi_m}\braket{E_0|\psi_m}^* dt_1 \cdots dt_m,$$ 
under which all terms with exponents that aren't identically 1 evaluate to 0. This only leaves the terms where all $a_i = b_i$ assuming non-degenerate states. For the degenerate case, the corresponding $\ket{E}$ vectors can be replaced by a sum of vectors that span the eigenbasis for that eigenvalue instead. The infinite time average success probability can then be simplified to
\begin{align}
P_{\infty} &= \sum_{a_1 \cdots a_m = 0}^{N-1}  \left|  \braket{E_0|E^{(m)}_{a_m}} \braket{E^{(m)}_{a_m} | E^{(m-1)}_{a_{m-1}}} \cdots \braket{E^{(1)}_{a_1}|\psi_0} \right|^2.
\end{align}
By expanding each term of the sum as a product of conjugates, it is possible to group up terms in such a way that the sums can be nested as follows,
\begin{align*}
P_{\infty} &= \sum_{a_m=0}^{N-1} \cdots \left( \sum_{a_2=0}^{N-1} \bra{E^{(2)}_{a_2}} \left( \sum_{a_1=0}^{N-1}  \braket{\psi_0 | E^{(1)}_{a_1}} \ket{E^{(1)}_{a_1}}
\bra{E^{(1)}_{a_1}} \braket{E^{(1)}_{a_1} | \psi_0} \right)  \ket{E^{(2)}_{a_2}} \ket{E^{(2)}_{a_2}}  \bra{E^{(2)}_{a_2}} \right) \cdots \braket{E_0|E^{(m)}_{a_m}} \braket{E^{(m)}_{a_m} | E_0}.
\end{align*}
The innermost sum is a matrix that is diagonal in the basis of $\hat{H}_1$, with the next innermost sum being diagonal in the basis of $\hat{H}_2$, and so on. In practice this can be calculated by computing the diagonal and then changing basis to the eigenbasis of the next Hamiltonian. This requires the eigendecomposition of all matrices and two matrix-matrix multiplications per stage. Since both these operations are $O(N^3)$ the computational complexity of this calculation is $O(mN^3)$, allowing exact simulation for small $n$. It is also possible to perform a partial decomposition using e.g the Lanczos algorithm to find only the lowest energy eigenvectors, but a formal error analysis of this approach proved difficult despite showing promise numerically.

\section{Statistics} \label{Stats}

It is possible to calculate some properties of the distribution of energy levels from the $J$ and $h$ parameters alone, for example the first few statistical moments of the distribution are written below. This has two main purposes, firstly it allows for exact calculation of the timescale in Equation \ref{time}, and secondly it can be used to estimate the spectral norm for normalisation of $\hat{H}_I$.
In this section, Ising problems in the form of Equation \ref{Ising} will be used. In the following formulae, $\odot$ denotes elementwise multiplication (i.e the Hadamard product) and summation is done over all matrix entries.
\begin{align*}
    \braket{\hat{H}_I} &= 0 \\
    \braket{\hat{H}_I^2} &= \frac{1}{2} \sum J \odot J \\
    \braket{\hat{H}_I^3} &= \sum J^2 \odot J\\
    \braket{\hat{H}_I^4} &= \frac{1}{16} \left(48 \left(\sum J^3 \odot J \right) + 12 \left(\sum J \odot J \right)^2 + 32 \left(\sum J \odot J \odot J \odot J\right) - 96 \left(\sum (J \odot J)^2 \right) \right)\\
    \braket{\hat{H}_I^5} &= \frac{1}{32} \left( 384 \left(\sum J^4 \odot J \right) + 160 \left(\sum J \odot J \right)\left(\sum J^2 \odot J \right) - 1920 \left(\sum (J^2 \odot J) (J \odot J)\right) + 1280 \left(\sum J^2 \odot J \odot J \odot J \right) \right).\\
\end{align*}
The derivation of these is as follows. Let a state $S$ with $n$ qubits have spins labelled as $S_i = \pm 1$ for $i \leq n$, then an average $\braket{\hat{H}_I}$ can be written using index notation as a sum over all states
$$\braket{\hat{H}_I} = \frac{1}{2N} \sum_{S} S_i S_j J_{ij}.$$
However, this sum will always be zero, as for a given state where some $S_i = 1$, there is a state where $S_i = -1$, cancelling out in the summation. The only exception to this is when $i = j$, but the diagonal of $J$ is all 0 by definition. Now consider the variance, 
\begin{align*}
    \braket{\hat{H}_I^2} &= \frac{1}{4N} \sum_{S} \left(S_i S_j J_{ij} \right)^2 \\
    &= \frac{1}{4N}\sum_{S} S_i S_j S_k S_l J_{ij} J_{kl}
\end{align*}
Once again, the only nonzero parts of this sum are where indices are paired up such that flipping a spin doesn't change the sign of the sum. Indices that appear in the same $J$ matrix can't be paired up either, as before. This time, there are two choices, $i = k, j = l$ and $i = l, j = k$, reducing the sum to
\begin{align*}
    \braket{\hat{H}_I^2} &= \frac{1}{2N} \sum_{S} S_i S_j S_i S_j J_{ij} J_{ij}\\
    &= \frac{1}{2N} \sum_{S} J_{ij} J_{ij} \\
    &= \frac{1}{2} J_{ij} J_{ij} \\
\end{align*}
This idea can be used to, in theory, calculate any statistical moment $\braket{\hat{H}_I^n}$. First, write down a sum over all states and $2n$ indices, then pair up those indices to find the parts of the sum that don't disappear when summing over all states.

Code to calculate these formulae has been provided in the github repository \cite{MultistageQW}. The method used is known not to be optimal, as the coefficients of each term are found via solving a linear system which is built from solving a number of small problems exactly. A purely combinatorial method should be possible, however the runtime of any algorithm grows exponentially for higher moments. This can be seen since for the $N^{th}$ moment, every integer partition of $N$ that doesn't contain a $1$ corresponds to a term. This means there are at least as many terms as the integer partition of $\lfloor \frac{N}{2} \rfloor$, which grows exponentially.

The sum over $\Delta_{j,k}^2$ from Section \ref{Timescales} can be derived using a similar method.  However, unlike the previous derivation the value of the $h$ parameters is important as remapping the problem to $n + 1$ qubits does change the result, so the Hamiltonian from Equation \eqref{IsingJ} will be used here. It simplifies the calculation to use a symmetric $J$, so use $J' = \frac{(J + J^T)}{2},$ where the sum in Equation \eqref{IsingJ} is now over all indices. Let a state be represented by $S_i = \pm 1$ as before, with $0 \leq i \leq n$, then the energy of that state is $E = S_i S_j J_{i,j} + h_i S_i$. Suppose $S_a$ is flipped. Then the change in energy is
\begin{align} \label{energydiff}
    \Delta_a &= -2S_i S_a J'_{i,a}  -2S_a S_i J'_{a,i} -2 h_a S_a\\
    &= -4 S_i S_a J'_{i,a} -2h_aS_a.
\end{align}
The change in energy squared is then
\begin{align*}
    \Delta_a^2 &= (-4 S_i S_a J'_{i,a} -2h_aS_a)^2 \\
    &= 16S_iS_jJ'_{i,a}J'_{j,a} + 16S_iJ'_{i,a}h_a + 4h_a^2 \\
\end{align*}
The sum over all choices of flipped spin is then
\begin{align*}
    \Delta^2 &= \sum_k 16S_iS_jJ'_{i,k}J'_{j,k} + 16S_iJ'_{i,k}h_k + 4h_k^2, \\
\end{align*}
and the average over all states is then
\begin{align*}
    \braket{\Delta^2} &= 16 \sum_{i,j} J'_{i,j} J'_{i,j} + 4 \sum_i h_i ^ 2 \\
    &= 8 \sum_{i,j} J_{i,j} J_{i,j} + 4 \sum_i h_i ^ 2
\end{align*}
which can be used to calculate a precise value of $t_s$.

\section{New Energy Spread heuristic} \label{heuristic}

For a random selection of 1000 qubit spin glass problems, the formula in Appendix \ref{Stats} gives a kurtosis of almost exactly 3, implying that the non-normal tails mentioned in \cite{Callison} is a small scale effect that disappears for larger problems.
This motivates a new heuristic, using the statistical information in the $J$ matrix to fit a probability distribution. A maximum entropy distribution is chosen as fitting this kind of distribution assumes no additional information about the true distribution. When the first $k$ statistical moments are used, this distribution will be $\exp(p(x))$, where $p(x)$ is a polynomial of order $k$. The difficulty of calculating this fit scales with $k$, and not with the problem size. The best available implementation seems to be PyMaxEnt \cite{PyMaxEnt}, however it can fail to converge. As such, a normal distribution is currently used for fitting, corresponding to $k = 2$.

Previous work has shown how these statistical moments can be calculated for a many-body system by sampling random states \cite{ManyBody}, which are then used to find a Chebyshev expansion of the density of states. We prefer the maximum entropy distribution as it guarantees non-negativity without assuming any additional information, which a Chebyshev expansion does not. 

Let the energy levels be treated as random variables drawn from a distribution with probability density $f(x) = \exp(p(x))$, and cumulative distribution $F(x) = \int_{-\infty}^x \exp(p(x)) dx$. A result from extreme value theory \cite{extreme} is that if
\begin{equation}
\label{extreme}
\lim_{x \rightarrow F^{-1}(1)} \frac{d}{dx} \frac{1 - F(x)}{f(x)} = 0,
\end{equation}
then the maximum value from the set of N energy levels follows a Gumbel distribution with parameters $\mu = F(1 - \frac{1}{N})$ and $\beta = \frac{1}{N f(\mu)}$.
Since $\frac{d}{dx} \exp(p(x)) = p'(x) \exp(p(x))$, then
\begin{align*}
    \frac{d}{dx} \frac{1 - F(x)}{f(x)} &= \frac{f(x)^2 - f'(x) (1 - F(x))}{f(x)^2} \\
    &= \frac{F(x) - 1}{f(x)} \frac{f'(x)}{f(x)} - 1 \\
    &= p'(x) \frac{\int_{-\infty}^x \exp(p(y)) dy - 1}{\exp(p(x))} - 1 \\
    &= p'(x) \frac{-\int_{x}^\infty \exp(p(y)) dy}{\exp(p(x))} - 1 .
\end{align*}
The integral here can be rewritten as an asymptotic expansion using integration by parts as follows,
\begin{align} \label{gumbel1}
    \int_{x}^\infty \exp(p(y)) dy &= \int_{x}^\infty \frac{1}{p'(y)} \frac{d}{dy}\exp(p(y)) dy \\ \label{gumbel2}
    &= \left[ \frac{1}{p'(y)} \exp(p(y))\right]^\infty_x + \int_{x}^\infty \frac{p''(y)}{p'(y)^2} \exp(p(y))dy\\ \
    &\rightarrow -\frac{1}{p'(x)} \exp(p(x)) \text{ as } x \rightarrow \infty .
\end{align}
If $p$ is of order $k > 0$, then $\frac{p''(x)}{p'(x)^2}$ is of order $-k$, so as $x \rightarrow \infty$ the integral on the second line is less than $x^{-k}$ times the original integral, which is a vanishing proportion. Substituting this result into Equation \eqref{extreme} shows it is satisfied, and so approximation by a Gumbel distribution is justified. The new heuristic for the spectral radius is therefore to let $f(x)$ be a normal distribution with mean $\braket{\hat{H}_I}$ (which is identically 0 for Ising problems) and variance $\braket{\hat{H}_I^2}$, then to calculate $\mu$ and $\beta$ for the resulting Gumbel distribution and return its mean $\mu + e_m \beta$, where $e_m$ is the Euler$-$Mascheroni constant.

There are multiple sources of error in this method. First, and suspected to be most significant, is the standard deviation of the Gumbel distribution, $\sigma = \frac{\pi}{\sqrt{6}} \beta$. Next is the error due to treating the energy levels like a maximum entropy distribution. This is difficult to quantify, but the skew and kurtosis can be calculated from Appendix \ref{Stats}, and then used in a Jarque-Bera normality test to see if there is a significant difference. This has been done for some random $J$ matrices of different types in \cite{MultistageQW}, with mixed results. Despite the skew and excess kurtosis approaching 0 for large problems, they do so slowly enough that the Jarque-Bera test shows significant deviations from normality. This is shown in Table \ref{tab:normality_failures}:
\begin{table}[h]
\centering
\begin{tabular}{|c|c|c|c|}
\hline
\textbf{n} & \textbf{Discrete Uniform} & \textbf{Uniform} & \textbf{Normal} \\ \hline
5  & 94   & 68   & 56   \\ \hline
10 & 513  & 292  & 282  \\ \hline
15 & 815  & 583  & 619  \\ \hline
20 & 1000 & 995  & 997  \\ \hline
\end{tabular}
\caption{Number of problems (out of 1000) failing the normality test at 5\% significance, for three methods of sampling $J$ parameter values.}
\label{tab:normality_failures}
\end{table}
The final source of error is how well the Gumbel distribution describes the extreme values for finite N. For smaller N it is possible to calculate the exact extreme value distribution. Let $F(x)$ be the CDF of the distribution of energy levels, and let $X_1, X_2, ..., X_N$ be N samples from it. Then
\begin{align*}
    P(\max(X_1, X_2, ..., X_N) < x) &= P(X_1 < y, X_2 < y, ..., X_N < x) \\
    &= P(X_1 < x) P(X_2 < x) ... P(X_N < x) \\
    &= F(x)^N
\end{align*}
is the exact CDF for the extreme value. Table \ref{tab:kl_values} shows the KL divergence between the Gumbel distribution and the true extreme value distribution for some values of $n$, showing that the difference is minimal and drops off exponentially.

\begin{table}[h]
\centering
\begin{tabular}{|c|c|}
\hline
\textbf{n} & \textbf{KL-Divergence} \\ \hline
5  & 0.00793 \\ \hline
10 & 0.00320 \\ \hline
15 & 0.00151 \\ \hline
20 & 0.00087 \\ \hline
25 & 0.00056\\ \hline
\end{tabular}
\caption{KL-divergence between Gumbel and true extreme value distribution for sample size $2^n$.}
\label{tab:kl_values}
\end{table}

\section{Analytic Example} \label{Example}

Grover search \cite{Grover} is a well known example of an analytically solvable quantum optimization, which is possible due to the quantum state being restricted to a 2D subspace throughout the algorithm. Likewise, by only considering $2 \times 2$ Hamiltonians, it is also possible to construct an operator which exactly has the ground state prescribed by Equation \eqref{spectralH}. The formula that results is not useful for larger dimensions, but doing so makes it clear that the Hamiltonians must be normalised by their energy spread.

The Pauli matrices and the identity form a basis for the vector space of $2 \times 2$ Hermitian matrices. Given any $2 \times 2$ operator, it can be shifted and rescaled into the form of a pure state density operator without changing the eigenvectors. Therefore, without loss of generality, let two operators $\hat{A}$ and $\hat{B}$ be defined by the Bloch vectors $\vec{a} = (a_x, a_y, a_z)$ and $\vec{b} = (b_x, b_y, b_z)$, so $\hat{A} = \frac{-1}{2}(\hat{I} + \vec{a} \cdot \boldsymbol{\hat{\sigma}})$ and $\hat{B} = \frac{-1}{2}(\hat{I} + \vec{b} \cdot \boldsymbol{\hat{\sigma}})$. A negative sign has been introduced so that the Bloch vectors correspond to the ground state and not the excited state. To simplify further, now rotate the Bloch sphere such that $\vec{a} = (0,0,1)$ and $b_y = 0$. The two ground states are then
$$ \ket{\psi_A} = \ket{0}$$
and
$$\ket{\psi_B} = \cos(\frac{\theta}{2})\ket{0} + \sin(\frac{\theta}{2})\ket{1}$$
Let $\nu \in \mathbb{R}$, the aim is to create an operator $\hat{A} + \gamma \hat{B}$ which has ground state $\ket{\psi_t} \propto \ket{\psi_A} + \nu \ket{\psi_B}$. Normalising gives
$$ \ket{\psi_t} = \frac{1}{\sqrt{1 + 2 \nu \cos(\frac{\theta}{2}) + \nu^2}} \left( (1 + \nu \cos(\frac{\theta}{2}))\ket{0} + \nu \sin(\frac{\theta}{2})\ket{1} \right),$$
which corresponds to the Bloch vector $\vec{t}$ with 
$$\theta_t = 2\tan^{-1} \left( \frac{\nu \sin(\frac{\theta}{2})}{1 + \nu \cos(\frac{\theta}{2})}\right).$$
$\vec{b} - b_z\vec{a}$ is orthogonal to \vec{a}, so these two vectors form an orthogonal basis. Expressing \vec{t} in this basis gives
\begin{align*}
    \vec{t} &= t_z \vec{a} + \frac{t_x}{b_x} (\vec{b} - b_z \vec{a}) \\
    &= (t_z - \frac{t_x}{b_x}b_z) \vec{a} + \frac{t_x}{b_x} \vec{b} \\
    &\propto \vec{a} + \frac{t_x}{b_x t_z - t_xb_z} \vec{b} \\
    &= \vec{a} + \frac{\sin{\theta_t}}{\cos{\theta_t} \sin{\theta} - \sin{\theta_t}\cos{\theta}} \vec{b} \\
    &= \vec{a} + \underbrace{\frac{\sin{\theta_t}}{\sin(\theta - \theta_t)}}_{\color{red}{\gamma}} \vec{b}.
\end{align*}
By the same shifting and scaling argument as before, the operator $\hat{T}$ corresponding to Bloch vector $\vec{t}$ has the same ground state as $\hat{A} + \gamma \hat{B}$, so this $\gamma$ gives the desired ground state.

In the case that $\theta = \pi$, this formula fails and gives $\vec{t} = \vec{a} + \vec{b} = \vec{0}$ no matter the value of $\nu$. In this situation, $\hat{A}$ and $\hat{B}$ are simultaneously diagonalisable, so $\hat{A} + \gamma \hat{B}$ has $\ket{\psi_A}$ as its ground state for $\gamma < 1$, $\ket{\psi_B}$ as its ground state for $\gamma > 1$, and for $\gamma = 1$ the operators add to the identity matrix, which has all vectors in its degenerate ground state.

\end{document}